\documentclass[onecolumn,showpacs,preprintnumbers,amsmath,amssymb,prb,floatfix]{revtex4}
\usepackage{graphicx}
\usepackage{dcolumn}
\usepackage{bm}

\begin{document}

\title{Interference through quantum dots}

\author{Y Tokura$^{1,2}$, H Nakano$^{1,3}$, and T Kubo$^2$}
\email{tokura@nttbrl.jp}
\affiliation{$^1$NTT Basic Research Laboratories, NTT Corporation, Atsugi-shi, Kanagawa 243-0198, Japan\\
$^2$Quantum Spin Information Project, ICORP, JST, Atsugi-shi, Kanagawa 243-0198, Japan\\
$^3$Department of Physics, Tokyo University of Science, Shinjuku-ku, 
Tokyo 162-8601, Japan}
\date{\today}

\begin{abstract}
We discuss the effect of quantum interference on transport through
a quantum dot system. 
We introduce an indirect coherent coupling parameter $\alpha$, 
which provides constructive/destructive interference
in the transport current depending on its phase and the magnetic flux.
We estimate the current through the quantum dot system
using the non-equilibrium Green's function method as well as the master equation
method in the sequential tunneling regime.
The visibility of the Aharonov-Bohm oscillation is evaluated.
For a large inter-dot Coulomb interaction, the current
is strongly suppressed by the quantum interference effect, while the current is restored
by applying an oscillating resonance field with the frequency of twice the inter-dot tunneling energy.
\end{abstract}

\maketitle

\section{Introduction}
Quantum phase coherence in mesoscopic systems is strikingly demonstrated
with the principle of superposition, or  interference experiments. 
Aharonov-Bohm (AB) interference is the most fundamental type and has been
experimentally confirmed in metallic and semiconductor rings.
Recently, in interference experiments with an Aharonov-Bohm 
ring containing a quantum dot (QD) in one of the arms, 
quasi-periodic modulation of the tunneling current has been
demonstrated as a function 
of the magnetic flux through the ring \cite{yacoby,schuster,ji}. 
This confirms that phase coherence is maintained during the tunneling process through a QD. 
The Fano effect is another type of interference in mesoscopic physics, which
occurs in a system in which discrete and continuum energy states coexist \cite{fe,kobayashi}.

More recently the AB oscillations of a tunneling current passing 
through a laterally coupled double quantum dot (DQD) system 
were observed \cite{holleitner,hatano}. 
These experimental results have motivated theoretical investigations 
of electron transport through such a system \cite{kubala,kang,bai}. 
DQD has been attracting attention as an important device structure 
for entangled spin qubit operations \cite{loss2,hatano2,petta}. 
There is also an interesting theoretical prediction that cotunneling 
currents passing through spin-singlet and triplet states have 
different AB oscillation phases \cite{loss}. 

In this paper, we consider the transport through an AB interferometer 
containing a laterally coupled DQD. 
We introduce the indirect coupling parameter $\alpha$, 
which characterizes the strength of the coupling 
via the reservoirs between two QDs \cite{shahbazyan}. 
A system with the maximum coupling $|\alpha|=1$ 
has already been widely studied theoretically \cite{kubala,kang,bai}. 
In actual systems, however, such a case is very special 
and most experimental situations correspond to $|\alpha|<1$. 
The situation where $\alpha=0$ has also been explored 
in the context of the orbital Kondo problem \cite{wilhelm,orbital}. 
We calculate the tunneling current through the DQD systems 
in terms of Green's function techniques for non-interacting systems \cite{MW,JWM} 
as well as the master equation method.
Although electron spin is crucial in the previous theoretical proposals, 
here we disregard it and focus on the quantum interference properties 
of  spinless electrons with/without inter-dot Coulomb 
interaction.

This paper is organized as follows. In Sec.~\ref{model}, 
a standard tunneling Hamiltonian is employed 
to describe an AB interferometer containing a laterally coupled DQD. 
We introduce the indirect coupling parameter $\alpha$. 
The current formula for non-interacting case is provided in Sec.~\ref{0} and
the visibility of the AB oscillation is discussed in the large bias limit.
In Sec.~\ref{U}, we provide the current expression in the limit of a strong inter-dot Coulomb interaction.
In some situations, the current is completely suppressed 
even when there is a large bias. 
Our results are summarized in Sec.~\ref{conclusion}. 
Three sections in the Appendices provides the detailed solutions of
the master equation.

\section{Model and formulation}\label{model}
\begin{figure}
\begin{center}
\includegraphics[width=0.48\textwidth]{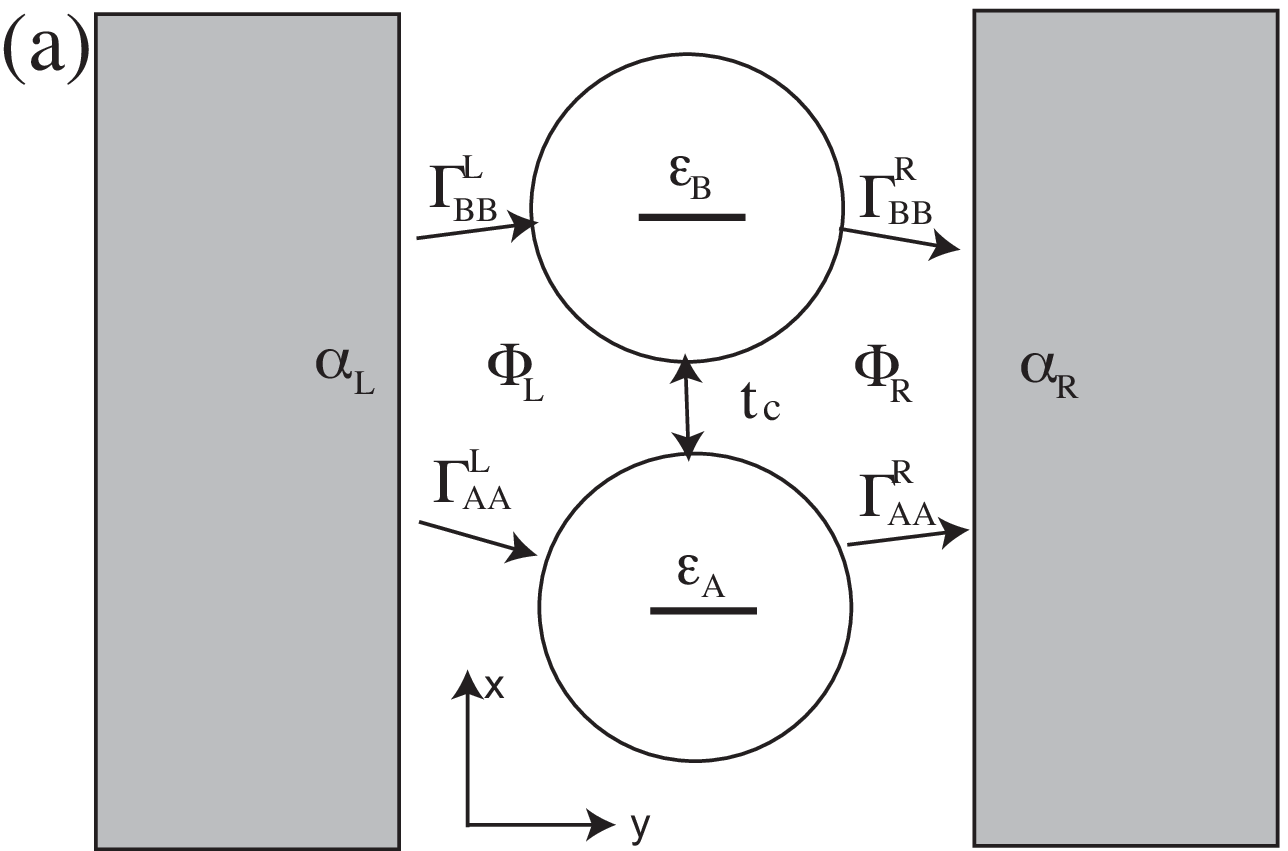}
\includegraphics[width=0.48\textwidth]{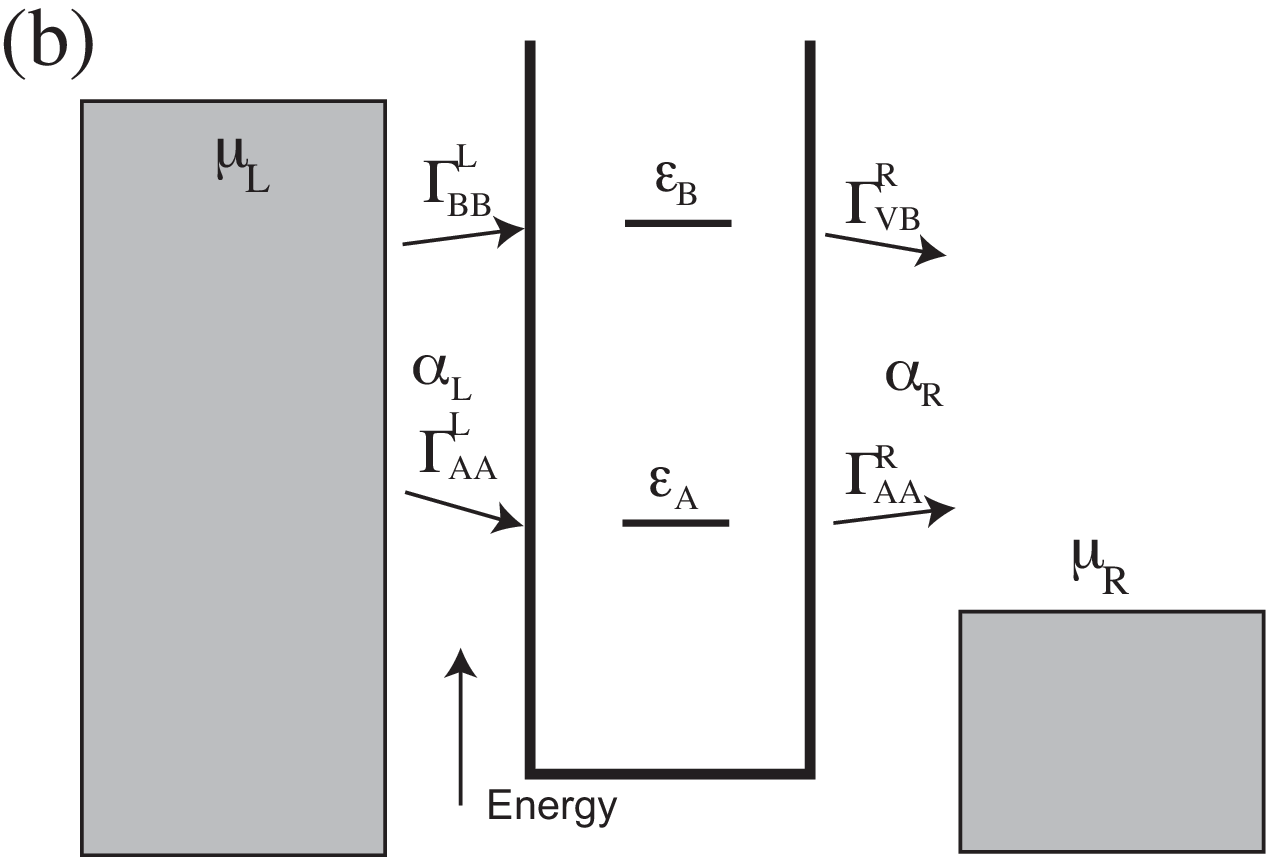}
\end{center}
\caption{\label{fig:system} (a) Schematic diagram of 
an AB interferometer containing a laterally coupled DQD. 
The magnetic fluxes threading the left and right sub-circuits are 
$\Phi_L$ and $\Phi_R$, respectively, and cause the AB effect. 
(b) Equivalent model of a single dot containing
two energy levels, which are in the bias window.}
\end{figure}

We studied laterally coupled double quantum dots (DQD) both of which are
tunnel-coupled to left (L) and right (R) reservoirs as shown in Fig.~(\ref{fig:system}a).
The Hamiltonian is ${\cal H}={\cal H}_R+{\cal H}_{DQD}+{\cal H}_T$ with
\begin{eqnarray}
{\cal H}_R&=&\sum_{\nu\in\{ L,R\}}\sum_k \epsilon_{\nu k}c_{\nu k}^\dagger c_{\nu k},\\
{\cal H}_{DQD}&=&\sum_{\zeta\in\{ A,B\}}\varepsilon_\zeta d_\zeta^\dagger d_\zeta
-t_c(d_A^\dagger d_B+h.c.)+U d_A^\dagger d_A d_B^\dagger d_B,\\
{\cal H}_T&=&\sum_{\nu\in\{L R\}}\sum_k\sum_{\zeta\in\{A,B\}}
[t_{\nu k}^{(\zeta)}(\phi_\nu)c_{\nu k}^\dagger d_\zeta+h.c.],
\end{eqnarray}
where $c_{\nu k}^\dagger (c_{\nu k})$ and $d_\zeta^\dagger (d_\zeta)$ represent 
creation (annihilation) operators of the reservoir $\nu=L/R$ and the quantum dot $\zeta=A/B$, respectively.
We disregarded the spin degree of freedom and we adopt a large limit for the
 intra-dot Coulomb interaction,
hence only one level is relevant in each dot.
$U$ and $t_c$ characterize the inter-dot Coulomb interaction and inter-dot tunneling amplitude, respectively.
We chose the gauge such that $t_c$ is real and positive.
$t_{L/R\ k}^{(A/B)}(\phi_{L/R})$ represents the tunneling amplitude between quantum dot
$A/B$ and the mode $k$ in the reservoir $L/R$.
The magnetic flux dependence of the tunneling amplitude is
$t_{\nu k}^{(B)}(\phi_\nu)/t_{\nu k}^{(A)}(\phi_\nu)=\exp(\mp i\phi_\nu)$, 
where  the upper (lower) sign is for $\nu=L(R)$, and the effective magnetic flux
$\phi_\nu=2\pi \Phi_\nu/\Phi_0$, which is defined by the flux $\Phi_\nu$ threading through the
area formed by DQD and the reservoir $\nu$ and the magnetic flux quantum $\Phi_0\equiv h/e$.
This Hamiltonian also describes the system of a single dot 
with {\it two} relevant energy levels as shown
in Fig.~(\ref{fig:system}b) when $t_c=0$ and $\phi_\nu=0$, 
where $U$ is now interpreted as an intra-dot Coulomb interaction.

In general, the tunneling current is obtained with the 
non-equilibrium Green's function (NEG) formalism by
\begin{eqnarray}\label{eq:keldysh}
I&=&\frac{ie}{2h}\int d\epsilon \mbox{Tr}\{(\bm{\Gamma}^L
-\bm{\Gamma}^R)\bm{G}^<(\epsilon)\nonumber\\
&&+[f_L(\epsilon)\bm{\Gamma}^L-f_R(\epsilon)\bm{\Gamma}^R]
[\bm{G}^r(\epsilon)-\bm{G}^a(\epsilon)]\},
\end{eqnarray}
where $\bm{G}^r(\epsilon)$ and $\bm{G}^a(\epsilon)$ are the retarded and advanced Green's
function of the DQD, and $\bm{G}^<(\epsilon)$ is the lesser Green's function \cite{MW,JWM}.
The boldface denotes the $2\times 2$ matrix and 
$f_\nu(\epsilon)\equiv 1/[1+e^{(\epsilon-\mu_\nu)/k_BT}]$ is the Fermi distribution function 
where $\mu_\nu, k_B$ and $T$ are the chemical potential of the reservoir $\nu$, 
the Bolzmann constant, and the absolute temperature, respectively.
The line-width functions are defined as
\begin{eqnarray}\label{eq:Gamma}
\Gamma_{\zeta \zeta'}^\nu(\epsilon)\equiv 2\pi\sum_k t_{\nu k}^{(\zeta)*}(\phi_\nu)
t_{\nu k}^{(\zeta')}(\phi_\nu)\delta(\epsilon-\epsilon_{\nu k}),
\end{eqnarray}
and the off-diagonal component has the following property
$\Gamma_{BA}^{\nu*}=\Gamma_{AB}^\nu$ for $\nu=L$ or $R$.
The wide-band limit approximation disregards  the energy
dependence of $\Gamma_{\zeta\zeta'}^\nu$.
The Green's functions have the following relations:
\begin{eqnarray}\label{eq:GF}
\bm{G}^a(\epsilon)=[\bm{G}^r(\epsilon)]^\dagger,
 \ \bm{G}^<(\epsilon)=\bm{G}^r(\epsilon)\bm{\Sigma}^<(\epsilon)\bm{G}^a(\epsilon),
\end{eqnarray}
where $\bm{\Sigma}^<(\epsilon)=i[f_L(\epsilon)\bm{\Gamma}^L
+f_R(\epsilon)\bm{\Gamma}^R]$ is the self-energy.
We have previously discussed the linear conductance for a zero offset 
$\Delta\equiv \varepsilon_B-\varepsilon_A=0$
in \cite{kubo} for non-interacting case ($U=0$), where the current formula is simpler as follows:
\begin{eqnarray}\label{eq:linear}
I&=&\frac{e}{h}\int d\epsilon\ [f_L(\epsilon)-f_R(\epsilon)]
\mbox{Tr}\{\bm{G}^r(\epsilon)\bm{\Gamma}^L\bm{G}^a(\epsilon)\bm{\Gamma}^R\}.
\end{eqnarray}
However, if the interaction $U$ is finite, we have to use Eq.~(\ref{eq:keldysh})
as demonstrated in the following section.

By contrast, we also derived a master equation using 
the method proposed by Gurvitz {\it et al.} \cite{gurvitz}, which is
appropriate for the large applied bias condition: 
$\mu_L-\mu_R\gg \Delta, \Gamma_{\zeta\zeta'}^\nu, t_c, k_B T$.
This approach can only handle the sequential tunneling process.
We specify the state of a DQD by the occupation number in these dots $(N_A,N_B)$.
The states $(0,0), (1,0), (0,1),(1,1)$ are abbreviated as $l=``0", ``A", ``B", ``D"$, respectively.
$\rho_{l,l'}$ is the reduced density matrix of a DQD after integrating out the 
reservoir modes in the total system density matrix.
The dynamics of the electrons passing through a DQD is characterized by the following
set of differential equations with an appropriate initial condition:
\begin{eqnarray}\label{master-eq:local}
\frac{d\rho_{00}}{dt}&=&
-(\Gamma_{AA}^L+\Gamma_{BB}^L)\rho_{00}
+\Gamma_{AA}^R\rho_{AA}
+\Gamma_{BB}^R\rho_{BB}
+\Gamma_{AB}^R\rho_{AB}
+\Gamma_{BA}^R\rho_{BA}\\\label{master-eq:2}
\frac{d\rho_{AA}}{dt}&=&
\Gamma_{AA}^L\rho_{00}
-(\tilde{\Gamma}_{BB}^L+\Gamma_{AA}^R)\rho_{AA}
+\tilde{\Gamma}_{BB}^R\rho_{DD}\\ \nonumber
&&+(\frac{\tilde{\Gamma}_{AB}^L-\Gamma_{AB}^R}{2}+i t_c)\rho_{AB}
+(\frac{\tilde{\Gamma}_{BA}^L-\Gamma_{BA}^R}{2}-i t_c)\rho_{BA}\\\label{master-eq:3}
\frac{d\rho_{BB}}{dt}&=&
\Gamma_{BB}^L\rho_{00}
-(\tilde{\Gamma}_{AA}^L+\Gamma_{BB}^R)\rho_{BB}
+\tilde{\Gamma}_{AA}^R\rho_{DD}\\ \nonumber
&&+(\frac{\tilde{\Gamma}_{AB}^L-\Gamma_{AB}^R}{2}-i t_c)\rho_{AB}
+(\frac{\tilde{\Gamma}_{BA}^L-\Gamma_{BA}^R}{2}+i t_c)\rho_{BA}\\\label{master-eq:4}
\frac{d\rho_{DD}}{dt}&=&
\tilde{\Gamma}_{BB}^L\rho_{AA}
+\tilde{\Gamma}_{AA}^L\rho_{BB}
-(\tilde{\Gamma}_{AA}^R+\tilde{\Gamma}_{BB}^R)\rho_{DD}-
\tilde{\Gamma}_{AB}^L\rho_{AB}-\tilde{\Gamma}_{BA}^R\rho_{BA}\\\label{master-eq:5}
\frac{d\rho_{AB}}{dt}&=&
\Gamma_{BA}^L\rho_{00}
+\frac{\tilde{\Gamma}_{BA}^L-\Gamma_{BA}^R}{2}(\rho_{AA}
+\rho_{BB})+i t_c(\rho_{AA}-\rho_{BB})\\ \nonumber
&&+(i \Delta-\frac{\tilde{\Gamma}_{AA}^L+\tilde{\Gamma}_{BB}^L
+\Gamma_{AA}^R+\Gamma_{BB}^R}{2})\rho_{AB}\\\label{master-eq:6}
\frac{d\rho_{BA}}{dt}&=&
\Gamma_{AB}^L\rho_{00}
+\frac{\tilde{\Gamma}_{AB}^L-\Gamma_{AB}^R}{2}(\rho_{AA}
+\rho_{BB})-i t_c(\rho_{AA}-\rho_{BB})\\ \nonumber
&&-(i \Delta+\frac{\tilde{\Gamma}_{AA}^L+\tilde{\Gamma}_{BB}^L
+\Gamma_{AA}^R+\Gamma_{BB}^R}{2})\rho_{BA}
\end{eqnarray}
The functions $\tilde{\Gamma}_{\zeta \zeta'}^\nu(\epsilon)$, which describe
the tunneling rate of electrons into (out of) dot(s) $\zeta,\zeta'$, when an electron
already occupying the DQD, are obtained by
replacing $\epsilon_{\nu k}$ in Eq.~(\ref{eq:Gamma}) with $\epsilon_{\nu k}-U$.
In the large limit for the interaction, $U \gg \mu_L-\mu_R$, 
the tunneling-in process $\tilde{\Gamma}_{\zeta \zeta'}^L$
is absent and we can set $\rho_{DD}=0$.

The above two approaches are sufficiently general for us to discuss the effect of
interaction and interference in the transport through a DQD. 
However, since the line-width function $\Gamma_{\zeta\zeta'}^\nu\ (\zeta\ne\zeta')$, which 
controls the strength of the coherence of the transport, is strongly 
dependent on the microscopic model, and we need further simplification 
to grasp the fundamental physics of this system.
Here, we define the indirect coherent coupling parameter $\alpha_\nu$, which was
 first introduced in \cite{shahbazyan}.
The explicit derivation of $\alpha_\nu$ is described in detail in \cite{kubo}.
Using $\alpha_\nu$, the off-diagonal part of the line-width function becomes
\begin{eqnarray}\label{eq:alpha}
\Gamma_{AB}^\nu&=&\alpha_\nu\sqrt{\Gamma_{AA}^\nu\Gamma_{BB}^\nu}e^{\mp i\phi_\nu},
\end{eqnarray}
where the upper (lower) sign is for $\nu=L(R)$.
All the parameters $\Gamma_{\zeta\zeta}^\nu$, $\alpha_\nu$ are independent of
energy in the wide-band limit.
We also disregarded the energy dependence of the effective flux induced by changes in
the electron trajectory.
The parameter $\alpha_L (\alpha_R)$ characterizes the coherent injection into the DQD from
the reservoir $L$ (the coherent emission from the DQD to the reservoir $R$). 
In general $\alpha_\nu$ is a complex parameter but the magnetic flux dependence 
is factorized in $\exp(\mp i\phi_\nu)$ as shown in Eq.~(\ref{eq:alpha}).
$|\alpha_\nu|=1$ corresponds to full coherence and $\alpha_\nu=0$ denotes 
zero coherence, corresponding to 
a situation where the two quantum dots are independently coupled to the reservoir.
For simplicity, we assume $\alpha_\nu$ is real and positive in the following argument.
There has been a detailed analysis of the coherence 
in a metallic reservoir  in \cite{nazarov}.

Equation~(\ref{master-eq:local}-\ref{master-eq:6}) differs from that  
obtained with a similar method \cite{jiang} and
from that obtained with the gradient expansion method \cite{dong}.
Both differ from ours as regards the sign of $\tilde{\Gamma}_{\zeta \zeta'}$ with
$\zeta\ne \zeta'$ and the former is missing the first term 
in Eqs.~(\ref{master-eq:5},\ref{master-eq:6})
which represents coherent injection of an electron into DQD from the left reservoir.
We can check that our formula provides a reasonable result in a limiting
situation as follows and in the next section.
Let us consider a symmetric system, namely, 
$\Gamma_{AA}^\nu=\Gamma_{BB}^\nu\equiv \gamma_\nu$, zero flux $\phi_\nu=0$, 
and non-interacting $U=0$.
We assume complete coherence $\alpha_R=\alpha_L=1$ and therefore
$\Gamma_{AB}^\nu=\Gamma_{BA}^\nu=\gamma_\nu$.
We transform from the dot A/B basis to the symmetric/antisymmetric (s/a) state
basis under the condition of zero offset $\Delta=0$.
The density matrix is then transformed to a new basis as
\begin{eqnarray}
\rho_{AA}&=&\frac{1}{2}(\rho_{ss}+\rho_{aa}+\rho_{sa}+\rho_{as}),\nonumber\\
\rho_{BB}&=&\frac{1}{2}(\rho_{ss}+\rho_{aa}-\rho_{sa}-\rho_{as}),\nonumber\\
\rho_{AB}&=&\frac{1}{2}(\rho_{ss}-\rho_{aa}-\rho_{sa}+\rho_{as}),\nonumber\\
\rho_{BA}&=&\frac{1}{2}(\rho_{ss}-\rho_{aa}+\rho_{sa}-\rho_{as}),\nonumber
\end{eqnarray}
with invariant $\rho_{00}$ and $\rho_{DD}$.
Then we define state dependent line-width functions: 
$\gamma_{s/a}^\nu=(1\pm\alpha_\nu)\gamma_\nu$
with $+(-)$ for a symmetric (antisymmetric) state.
The master equation for the new basis is
\begin{eqnarray}\label{master-eq:s-a1}
\frac{d\rho_{00}}{dt}&=&-(\gamma_s^L+\gamma_a^L)\rho_{00}
+\gamma_s^R\rho_{ss}+\gamma_a^R\rho_{aa},\\\label{master-eq:s-a2}
\frac{d\rho_{ss}}{dt}&=&-(\gamma_s^R+\gamma_a^L)\rho_{ss}
+\gamma_s^L\rho_{00}+\gamma_a^R\rho_{DD},\\\label{master-eq:s-a3}
\frac{d\rho_{aa}}{dt}&=&-(\gamma_s^L+\gamma_a^R)\rho_{aa}
+\gamma_a^L\rho_{00}+\gamma_s^R\rho_{DD},\\\label{master-eq:s-a4}
\frac{d\rho_{DD}}{dt}&=&-(\gamma_s^R+\gamma_a^R)\rho_{DD}
+\gamma_a^L\rho_{ss}+\gamma_s^L\rho_{aa},\\\label{master-eq:s-a5}
\frac{d\rho_{sa}}{dt}&=&-[\frac{1}{2}(\gamma_s^L+\gamma_a^L
+\gamma_s^R+\gamma_a^R)
+2it_c]\rho_{sa},
\end{eqnarray}
and there is a similar equation for $\rho_{as}$.
Because of the relaxation term $(\gamma_s^L+\gamma_a^L+\gamma_s^R+\gamma_a^R)/2$,
the quantum coherence term $\rho_{sa}$ simply disappears from any initial condition
for the steady state limit.
Therefore, Eqs.~(\ref{master-eq:s-a1}-\ref{master-eq:s-a5}) 
correctly describe the independent dynamics of symmetric and antisymmetric
channels with state dependent line-width functions.
It should be noted that when $\alpha_\nu=1$, the antisymmetric state 
is decoupled from the reservoir $\nu$,  $\gamma_a^\nu=0$.
This is because of the perfect destructive interference.

From Eqs.~(\ref{master-eq:local}-\ref{master-eq:6}), 
we obtain the steady state density matrix at $t\rightarrow \infty$
by employing the auxiliary relation $\rho_{00}+\rho_{AA}+\rho_{BB}+\rho_{DD}=1$.
Using the result, the steady current is obtained as follows:
\begin{eqnarray}\label{eq:current-master}
\frac{I}{e}&=&\Gamma_{AA}^R\rho_{AA}+\Gamma_{BB}^R\rho_{BB}
+(\tilde{\Gamma}_{AA}^R+\tilde{\Gamma}_{BB}^R)\rho_{DD}
+\Gamma_{AB}^R\rho_{AB}+\Gamma_{BA}^R\rho_{BA}.
\end{eqnarray}

\section{Noninteracting system}\label{0}

First we discuss the system without interaction ($U=0$).
For simplicity, we restrict to the symmetric coupling situation,
$\Gamma_{\mu\mu}^\nu\equiv\gamma$ and
symmetric fluxes $\phi_L=\phi_R\equiv\phi/2$.
The retarded Green's function is
\begin{eqnarray}
\bm{G}^r(\epsilon)&=&
\left(
  \begin{array}{cc}
  \epsilon-\varepsilon_A+i\gamma & 
  t_c+\frac{i}{2}\gamma(\alpha_L e^{-i\frac{\phi}{2}}+\alpha_R e^{i\frac{\phi}{2}}) \\
  t_c+\frac{i}{2}\gamma(\alpha_Le^{i\frac{\phi}{2}}+\alpha_Re^{-i\frac{\phi}{2}})
  & \epsilon-\varepsilon_B+i\gamma
   \end{array}
\right)^{-1}.
\end{eqnarray}
For noninteracting conductor, the transmission probability of the electron with
energy $\epsilon$ is defined as
\begin{eqnarray}
{\cal T}(\epsilon)&=&\mbox{Tr}\{\bm{G}^r(\epsilon)\bm{\Gamma}^L
\bm{G}^a(\epsilon)\bm{\Gamma}^R\},
\end{eqnarray}
which appears in Eq.~(\ref{eq:linear}).
The linear conductance $G$ at zero temperature is obtained in the Landauer formula
\begin{eqnarray}\label{eq:landauer}
G=\frac{e^2}{h}{\cal T}(0),
\end{eqnarray}
where the energy is measured from the (average) chemical potential of the reservoirs.
The explicit formula for zero-offset, $\Delta=0$, is shown in Eq.(18) of 
Ref.\cite{kubo}. (The definition of the sign of $t_c$ is reversed.)
The function ${\cal T}(0)$ for $\alpha=0$ and $\alpha=1$ at zero flux $\phi=0$ has
following simple physical meaning:
\begin{eqnarray}\label{eq:t0}
{\cal T}(0)=\left \{
\begin{array}{ccc}
  \frac{\gamma^2}{(\epsilon_0-t_c)^2+\gamma^2}
  +\frac{\gamma^2}{(\epsilon_0+t_c)^2+\gamma^2} &\mbox{for}& \alpha=0,\\
  \frac{(2\gamma)^2}{(\epsilon_0-t_c)^2+(2\gamma)^2} &\mbox{for}& \alpha=1,
   \end{array} \right.
\end{eqnarray}
where $\alpha=0$ corresponds to two independent Breit-Wigner resonances through
the symmetric and antisymmetric states with line-width $\gamma$, 
while $\alpha=1$ represents Breit-Wigner resonance through
only the symmetric state with doubled line-width $2\gamma$.
It had been shown that the period of AB oscillation of the linear
conductance is $2\pi$ when $t_c=0$ and
$4\pi$ when $t_c\ne 0$.
For non-integer flux, the $\epsilon_0$ dependence of the conductance 
shows Fano line shape when $\alpha=1$ \cite{fe}.
However, this Fano effect is quickly suppressed if $\alpha$ becomes less than 1.

\begin{figure}
\begin{center}
\includegraphics[width=\textwidth]{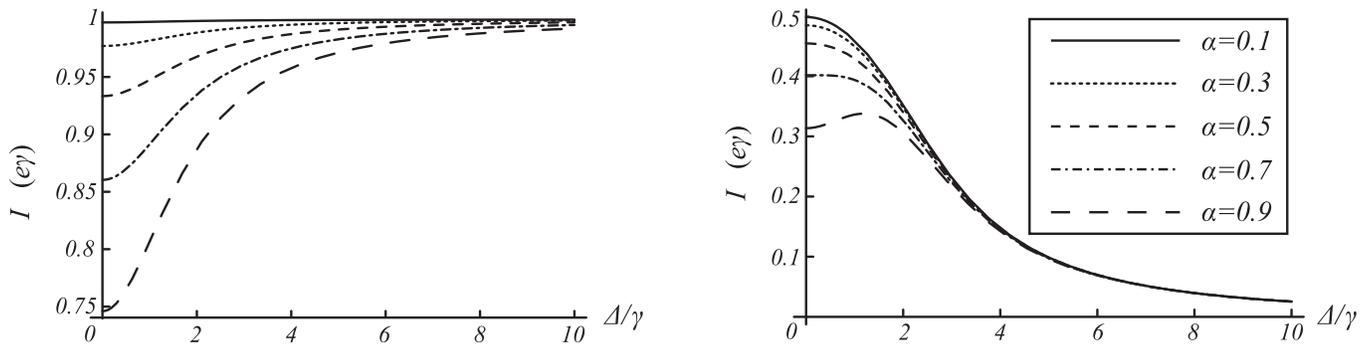}
\end{center}
\caption{\label{fig:non-int} $\Delta$ dependence of the current  with various
coherent coupling parameter $\alpha=\alpha_R$ for left: infinite bias and for right: 
$\mu_L-\mu_R=2\gamma$. $t_c=\epsilon_0=0, \alpha_L=0$ and $\phi=0$.}
\end{figure}

The current for a finite bias $eV\equiv \mu_L-\mu_R$ at $T=0$ is
\begin{eqnarray}
I&=&\frac{e}{h}\int_{\mu_R}^{\mu_L}\ d\epsilon {\cal T}(\epsilon).
\end{eqnarray}
In the limit of a large bias, this can be evaluated by the contour integral 
and the result for $\alpha_L=\alpha_R\equiv \alpha$ and $\Delta=0$ is
\begin{eqnarray}\label{eq:cur-non-linear}
I&=&e\gamma \frac{t_c^2+\gamma^2(1-\alpha^2\sin^2\frac{\phi}{2})}{t_c^2+\gamma^2}.
\end{eqnarray}
Now the period of the current oscillation with the flux is $2\pi$ independent of $t_c$.
At zero flux, $\phi=0$, the current is $e\gamma$ independent of $\alpha$, 
which is explicitly checked from Eq.~(\ref{eq:t0}) 
by replacing $\epsilon_0$ with $\epsilon_0-\epsilon$
and by integrating with $\epsilon$ for $(-\infty,\infty)$.
The current is the sum of  $e\gamma/2$ from symmetric and antisymmetric
states for $\alpha=0$, 
and the current is $e(2\gamma)/2$ from symmetric state for $\alpha=1$.
The energy offset $\Delta$ dependence of the current is shown 
for various values of $\alpha$ in Fig.~\ref{fig:non-int}.
It should be noted that for sufficiently large offset $\Delta\gg \gamma$,
the current is independent of $\alpha$.

The current for the large bias limit is also
derived by the master equation with $\tilde{\Gamma}_{\zeta\zeta'}^\nu=\Gamma_{\zeta\zeta'}^\nu$.
The result is shown in Eq.~(\ref{app-eq:non-int}) in \ref{sec:app1}.
For $\alpha_R=\alpha_L\equiv \alpha$ and $\phi_L=\phi_R=\phi/2$, we have
\begin{eqnarray}
I&=&e\gamma\frac{\Delta^2
+4(1-\alpha^2\cos^2\frac{\phi}{2})(t_c^2+\gamma^2(1-\alpha^2\sin^2\frac{\phi}{2}))}
{\Delta^2+4(1-\alpha^2\cos^2\frac{\phi}{2})(t_c^2+\gamma^2)},
\end{eqnarray}
which provides the same result for $\Delta=0$ as that obtained 
by the NEG method, Eq.~(\ref{eq:cur-non-linear}).
The current is the maximum, $I=e\gamma$,
at $\phi=2n\pi$ and the minimum at $\phi=(2n+1)\pi$ with an integer $n$.
The visibility of the AB oscillation is
\begin{eqnarray}
v&\equiv&\frac{I(\phi=0)-I(\phi=\pi)}{I(\phi=0)}=
\frac{4\gamma^2\alpha^2}{\Delta^2+4(t_c^2+\gamma^2)},
\end{eqnarray}

As shown in Sec.~\ref{model}, the Hamiltonian also describes
the system of a single dot with two levels.
Usually, coherent injection process to the multiple levels in a quantum dot
does not considered.
Here we clarify the condition when this is justified.
By putting $t_c=0$ and $\phi_\nu=0$ in Eq.~(\ref{app-eq:non-int}), we have
the current formula:
\begin{eqnarray}
I&=&e\gamma\frac{\Delta^2+\gamma^2\{4-2(\alpha_L^2+\alpha_R^2)\}}
{\Delta^2+\gamma^2\{4-(\alpha_L+\alpha_R)^2\}}.
\end{eqnarray}
The total current deviates from $e\gamma$, just the sum of the current via 
each level, $\frac{1}{2}e\gamma$, by the effect of quantum interference
when $\alpha_L\ne \alpha_R$.
This effect is maximum if $\Delta=0$ and one of the $\alpha$'s is one and the other
is zero,  where the current becomes $\frac{2}{3}e\gamma$.
This effect of interference vanishes for large offset $|\Delta|\gg \gamma$ and the current
is $e\gamma$.
This behavior is shown in Fig.~\ref{non-int}(left).

\section{Strong interaction limit}\label{U}

Here we consider the case of $U\rightarrow \infty$.
The general form of the steady current is obtained in \ref{sec:app3}.
For simplicity, we restricted ourselves to the symmetrical coupling 
$\Gamma_{AA}^\nu=\Gamma_{BB}^\nu\equiv\gamma_\nu$.
In the special case where zero flux $\phi_\nu=0$,  we obtain from Eq.~(\ref{eq:cur-uinf}) 
\begin{eqnarray}\label{eq:zeroflux}
I&=&e\frac{2\gamma_R\gamma_L}{2\gamma_L+\gamma_R}
\frac{\frac{\Delta^2}{\gamma_R^2+4t_c^2}+1-\alpha_R^2}
{\frac{\Delta^2}{\gamma_R^2+4t_c^2}+\frac{2\gamma_L+(1-\alpha_R^2)\gamma_R
-2\alpha_L\alpha_R\gamma_L}{2\gamma_L+\gamma_R}}.
\end{eqnarray}
When the electron tunneling-out process to the reservoir $R$ is incoherent, $\alpha_R=0$, 
the current value becomes the classical  limit, 
\begin{eqnarray}
I_{incoherent}&=&e\frac{2\gamma_R\gamma_L}{2\gamma_L+\gamma_R}.
\end{eqnarray}
Interestingly, if $\alpha_L=\alpha_R$, the current has the same
value $I_{incoherent}$ as if the coherent transport is absent.
In both cases, the current value is independent of $\Delta$ and $t_c$.
Under general $\alpha_\nu$ conditions, the current value approaches 
$I_{incoherent}$ if the condition $\Delta^2\gg \gamma_R^2+4t_c^2$ is satisfied.
Similar formula is derived for more general situation (asymmetric couplings) 
in Eq.~(\ref{eq:cur-single}).

\begin{figure}
\begin{center}
\includegraphics[width=0.49\textwidth]{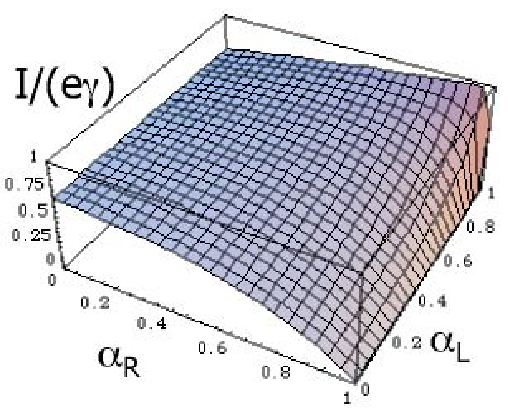}
\includegraphics[width=0.49\textwidth]{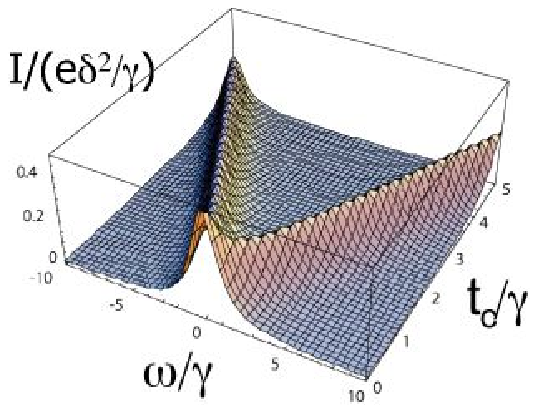}
\end{center}
\caption{\label{fig:CPT} (a) Left, the $\alpha_\nu$ dependence of the current 
in the limit of large $U$ and $\Delta=0$. The current is strongly suppressed near
$\alpha_R=1$. (b) Right, photon induced leakage current caused by applying weak oscillating field of
frequency $\omega$. The DQD is in the current suppressed condition: $\alpha_R=1, \Delta=0$.}
\end{figure}

When $\alpha_R \rightarrow 1$  and $\Delta\rightarrow 0$, 
the current is completely suppressed even if we supply the system with a large
bias. We need to keep $\alpha_L<1$, since $\alpha_R=\alpha_L=1$
provides finite current $I_{incoherent}$.
This is evident from the plot of overall dependence on $\alpha_R$ and $\alpha_L$ in
Fig.~\ref{fig:CPT}(a).
This can be understood as the system being trapped in the  `dark-state',
which in this context means the antisymmetric state that cannot couple to the
$R$ reservoir as discussed in the previous section.
The steady state density matrix in this limit is
\begin{eqnarray}
\rho_{AA}&=&\rho_{BB}=\frac{1}{2},\ \rho_{AB}=\rho_{BA}=-\frac{1}{2},\ \rho_{00}=0,\nonumber
\end{eqnarray}
which is the density matrix of the pure state:
$\rho=|\Psi_a\rangle\langle \Psi_a|$ with the
antisymmetric state $|\Psi_a\rangle\equiv \frac{1}{\sqrt{2}}(|A\rangle-|B\rangle)$.
This mechanism has been discussed in a triple dot system as a
coherent population trapping (CPT) mechanism \cite{michaelis}. 
The current in such a system is estimated in \ref{sec:app2} and 
we found the $\Delta$ dependence is similar to Eq.~(\ref{eq:zeroflux}).

We demonstrate the collapse of current suppression by applying 
an oscillating electric field \cite{stoof,hazelzet,dong}.
We evaluated the effect of a weak oscillating field $\Delta(t)=\delta \cos \omega t$ for $\alpha_R=1$,
and found the leakage current in the lowest perturbation in $\delta$,
\begin{eqnarray}
I|_{\alpha_R=1}&=&\frac{e\gamma_R\delta^2(\gamma_R^2+4t_c^2+\omega^2)}
{2(1-\alpha_L)\{\gamma_R^4+(4t_c^2-\omega^2)^2+2\gamma_R^2(4t_c^2+\omega^2)\}},
\end{eqnarray}
which is plotted in Fig. ~\ref{fig:CPT}(b).
The current peaks with a value 
$\frac{e\delta^2}{2\gamma_R(1-\alpha_L)}\frac{\gamma_R^2+8t_c^2}{\gamma_R^2+16t_c^2}$
at a frequency $\omega\sim \pm 2 t_c$, which corresponds to the emission of
one photon and the system transits from the antisymmetric state to the symmetric state,
that allows the electron to leak in the right reservoir.

The flux dependence of the current in the large bias limit is following.
We only consider the symmetric configuration:
$\gamma_R=\gamma_L=\gamma$ and $\phi_L=\phi_R=\phi/2$ and $\alpha_R=1$
and use Eq.~(\ref{eq:cur-uinf}).
When the offset $\Delta$ is zero,
\begin{eqnarray}
I&=&e\gamma\frac{2t_c^2(1-\cos\phi)}{\gamma^2+(5-2\alpha_L)t_c^2-(\alpha_L\gamma^2+(1+2\alpha_L)t_c^2)\cos\phi},
\end{eqnarray}
therefore the visibility of AB oscillation of the current is 1.
When the inter-dot tunneling $t_c$ is zero,
\begin{eqnarray}\label{eq:cur-ab-delta}
I&=&e\gamma\frac{2\Delta^2}{3\Delta^2+2\alpha_L\Delta\gamma\sin\phi
+2\gamma^2(1-\alpha_L\cos\phi)},
\end{eqnarray}
where its visibility is
\begin{eqnarray}
v&=&\frac{4\alpha_L\gamma\sqrt{\gamma^2+\Delta^2}}
{3\Delta^2+2\gamma^2+2\alpha_L\gamma\sqrt{\gamma^2+\Delta^2}},
\end{eqnarray}
which is a monotonic decreasing function of $|\Delta|/\gamma$ and $\alpha_L$.
The linear (two-terminal) conductance should be symmetric with respect to the
reversal of the flux in accordance with Onsager's relation, hence is an even function of $\phi$.
However, the current shown Eq.~(\ref{eq:cur-ab-delta}) is obviously not an even 
function. 
This is not the problem since we are discussing the current in a non-linear regime,
that is out of the boundary of Onsager's argument about a linear-response regime.

We also studied the current expression using the NEG method 
for the special condition $\Delta=0$.
For simplicity, all tunneling amplitudes are the same and we use the single
parameter $\gamma$. 
We transform the basis to the symmetric/antisymmetric basis
with $\epsilon_s=\epsilon_0-t_c$ and $\epsilon_a\equiv \epsilon_0+t_c$ and
$\epsilon_0=(\varepsilon_A+\varepsilon_B)/2$.
The inter-dot interaction Hamiltonian becomes $U n_sn_a$
where $n_n\equiv d_n^\dagger d_n$ with $n=s/a$.
On this basis, the line-width function matrix is
\begin{eqnarray}
\bm{\Gamma}^\nu=\gamma
\left (
  \begin{array}{cc}
    1+\alpha_\nu   & 0  \\
    0  &  1-\alpha_\nu  
  \end{array}
\right ).
\end{eqnarray}
We define the total line-width $\bm{\Gamma}\equiv \bm{\Gamma}^L+\bm{\Gamma}^R$.
We calculated the retarded Green's function 
$G_{nm}^r(t-t')=-i\theta(t-t')\langle\{d_n(t),d_{m}^\dagger(t')\}\rangle$, where $n,m=s/a$ using the equation of motion (EOM) approach combined with the lowest order decoupling approximation,
which is after Fourier transform
\begin{eqnarray}
\bm{G}_{nm}^r(\epsilon)&=&\frac{\delta_{nm}\chi(\epsilon)}
{\epsilon-\epsilon_n+\frac{i}{2}\Gamma_{nn}\chi(\epsilon)},
\end{eqnarray}
where $\chi(\epsilon)\equiv 1+\frac{U\langle n_{\bar{n}}\rangle}{\epsilon-\epsilon_n-U}$.

Then we have in the limit of $U\rightarrow \infty$,
\begin{eqnarray}
G_{nn}^r(\epsilon)&=&\frac{1-\langle n_{\bar{n}}\rangle}{\epsilon-\epsilon_n
+\frac{i}{2}\Gamma_{nn}(1-\langle n_{\bar{n}}\rangle)},
\end{eqnarray}
and the lesser Green's function is derived from Eq.~(\ref{eq:GF}).
Finally, the populations of the states are given by
\begin{eqnarray}\label{eq:population}
\langle n_n\rangle&=&\int \frac{d\epsilon}{2\pi i}G_{nn}^<(\epsilon),
\end{eqnarray}
which should be evaluated self-consistently.
The linear conductance at zero temperature for $\alpha_L=\alpha_R=\alpha$ is obtained by
\begin{eqnarray}\label{eq:lowest}
G&=&\frac{e^2}{h}\{\frac{\tilde{\gamma}_s^2}{\epsilon_s^2+\tilde{\gamma}_s^2}
+\frac{\tilde{\gamma}_a^2}{\epsilon_a^2+\tilde{\gamma}_a^2}\nonumber \\
&&-2(\alpha\sin(\frac{\phi}{2}))^2(1-\langle n_s\rangle)(1-\langle n_a\rangle)
\frac{\epsilon_s\epsilon_a+\tilde{\gamma}_s\tilde{\gamma}_a}
{(\epsilon_s^2+\tilde{\gamma}_s^2)
(\epsilon_a^2+\tilde{\gamma}_a^2)}\},
\end{eqnarray}
where $\tilde{\gamma}_{s/a}\equiv\gamma(1\pm\alpha\cos(\frac{\phi}{2}))(1-\langle n_{a/s}\rangle)$.

In the limit of a large bias, Eq.~(\ref{eq:population}) reduces to the following self-consistent equation 
\begin{eqnarray}
\langle n_{s/a}\rangle &=&\frac{(1\pm\alpha_L)(1-\langle n_{a/s}\rangle)}{2\pm\alpha_L\pm\alpha_R},
\end{eqnarray}
and the solutions are
\begin{eqnarray}
\langle n_{s/a}\rangle &=&\frac{(1\pm\alpha_L)(1\mp\alpha_R)}{3-2\alpha_L\alpha_R-(\alpha_R)^2}.\end{eqnarray}
Putting these in the formula of the current,
\begin{eqnarray}
I&=&\sum_{n\in\{s, a\}}I_n,\\
I_{s/a}&=&e\gamma (1\pm\alpha_L)(1\mp\alpha_R)
\frac{1-\langle n_{a/s}\rangle}{2\pm\alpha_L\pm\alpha_R},
\end{eqnarray}
we finally obtain the same result as that obtained by the master equation, Eq.~(\ref{eq:zeroflux}) 
with $\Delta=0$:
\begin{eqnarray}
I&=&2e\gamma\frac{1-\alpha_R^2}{3-2\alpha_L\alpha_R-\alpha_R^2}.
\end{eqnarray}

\section{Conclusions}\label{conclusion}
We discussed the effect of quantum interference on the transport through
a quantum dot system. We stressed  the role of the indirect coherent
coupling parameter $\alpha$, which provides constructive/destructive interference
in the transport current depending on its phase.
We derived the current using the non-equilibrium Green's function method
 as well as the master equation
method in the sequential tunneling regime.
For a large inter-dot Coulomb interaction,  the current
is strongly suppressed by the quantum interference effect, where the current is restored
by applying oscillating resonant field.

\begin{acknowledgements}
We thank A. Aharony, O. Entin-Wohlman, J. Tobiska, M. Pioro-Ladri\`{e}re, T. Hatano, 
and S. Tarucha for valuable discussions and useful comments. 
One of the authors (Y. T.) is partly supported by SORST-JST.
\end{acknowledgements}

\appendix
\section{Derivation of current for non-interacting system}\label{sec:app1}
Steady state density matrix elements are derived
from the master equation by setting $\frac{d\rho_{ll'}}{dt}=0$
in Eqs.~(\ref{master-eq:local}-\ref{master-eq:6}).
When the interaction is absent, $\tilde{\Gamma}_{\zeta\zeta'}^\nu=\Gamma_{\zeta\zeta'}^\nu$.
Since the six algebraic equations are not independent, we need
the equation for conservation of probability, 
$\rho_{00}+\rho_{AA}+\rho_{BB}+\rho_{DD}=1$.
Evaluating the current with Eq.~(\ref{eq:current-master}), we obtain
\begin{eqnarray}\label{app-eq:non-int}
I&=&e\gamma \frac{N_I}{D_I},\\
N_I&=&\Delta^2+\gamma^2[4-2(\alpha_L^2+\alpha_R^2)
+\alpha_L^2\alpha_R^2\sin^2(\phi_L+\phi_R)]\nonumber\\
&+&2t_c^2[2-\alpha_L^2\cos^2\phi_L-\alpha_R^2\cos^2\phi_R],\nonumber \\
D_I&=&\Delta^2+\gamma^2[4-\alpha_L^2-\alpha_R^2-2\alpha_L\alpha_R\cos(\phi_L+\phi_R)]\nonumber\\
&+&t_c^2[4-(\alpha_L\cos\phi_L+\alpha_R\cos\phi_R)^2].\nonumber 
\end{eqnarray}

\section{Derivation of current in the strong inter-dot interaction limit}\label{sec:app3}
We solve the master equation Eqs.~(\ref{master-eq:local}-\ref{master-eq:6}) 
for $U\rightarrow \infty$, namely, neglecting the term $\rho_{DD}$ 
and $\tilde{\Gamma}_{\zeta\zeta'}^\nu$ in the steady state condition.
First, we restrict ourselves to the symmetrical coupling 
$\Gamma_{AA}^\nu=\Gamma_{BB}^\nu\equiv\gamma_\nu$ and the result is
\begin{eqnarray}\label{eq:cur-uinf}
I&=&e\frac{2\gamma_R\gamma_L}{D}\{\Delta^2
-2\alpha_R^2t_c^2\cos2\phi_R+(1-\alpha_R^2)\gamma_R^2+2(2-\alpha_R^2)t_c^2\},\\
D&=&\Delta^2(2\gamma_L+\gamma_R)+2\alpha_L\alpha_R\Delta \gamma_L\gamma_R
\sin(\phi_L+\phi_R)+(1-\alpha_R^2)\gamma_R^{3}\nonumber\\
&&-2\alpha_L\alpha_R\gamma_L(\gamma_R^{2}+2t_c^2)\cos(\phi_L+\phi_R)
-4\alpha_L\alpha_R\gamma_L t_c^2\cos(\phi_R-\phi_L)\nonumber\\
&&-2\alpha_R^2\gamma_Rt_c^2\cos2\phi_R+2(2-\alpha_R^2)\gamma_Rt_c^2
+2\gamma_L(\gamma_R^{2}+4t_c^2).
\end{eqnarray}

In the model of a single quantum dot with two levels ($t_c=0, \phi_\nu=0$)
as shown in Fig.~\ref{fig:system}(b), current is evaluated
for the most general choices of $\Gamma_{\zeta\zeta'}^\nu$,
\begin{eqnarray}\label{eq:cur-single}
I&=&I_{incoherent}K(\Delta),\\
I_{incoherent}&=&e\frac{(\Gamma_{AA}^L+\Gamma_{BB}^L)\Gamma_{AA}^R\Gamma_{BB}^R}
{\Gamma_{AA}^R\Gamma_{BB}^R+\Gamma_{AA}^R\Gamma_{BB}^L
+\Gamma_{BB}^R\Gamma_{AA}^L},\\
K(\Delta)&=&\frac{4\Delta^2+(1-\alpha_R^2)(\Gamma_{AA}^R+\Gamma_{BB}^R)^2}
{4\Delta^2+
\frac{(1-\alpha_R^2)\Gamma_{AA}^R\Gamma_{BB}^R
+\Gamma_{AA}^R\Gamma_{BB}^L+\Gamma_{BB}^R\Gamma_{AA}^L
-2\alpha_L\alpha_R\kappa}
{\Gamma_{AA}^R\Gamma_{BB}^R+\Gamma_{AA}^R\Gamma_{BB}^L
+\Gamma_{BB}^R\Gamma_{AA}^L}(\Gamma_{AA}^R+\Gamma_{BB}^R)^2},\nonumber
\end{eqnarray}
where $\kappa\equiv \sqrt{\Gamma_{AA}^R\Gamma_{BB}^R\Gamma_{AA}^L\Gamma_{BB}^L}$.
In this model, the incoherent current is suppressed if one of the coupling to the
right reservoir, $\Gamma_{AA}^R$ or $\Gamma_{BB}^R$, is very small \cite{belzig}.
The corresponding local state in the quantum dot is now the 'dark state' to suppress
the current.
The function $K(\Delta)$ is 1 when the coherent coupling in the right reservoir is absent,
$\alpha_R=0$ irrespective of the values of $\Delta, \alpha_L,$ and so forth.
When $\alpha_R\rightarrow 1$, $K(\Delta)$ is suppressed for small $\Delta$, while
for large offset, $\Delta$, $K(\Delta)\rightarrow 1$.

\section{Derivation of current through the triple dot system}\label{sec:app2}
We consider the triple-dot model used in Ref.~\cite{michaelis} with two lateral dots, A and B,
with energies $\varepsilon_A$ and $\varepsilon_B$ coupled to left reservoirs
independently and one dot, C, with energy $\varepsilon_c$ coupled to right reservoir.
The dot $A$ and $B$ are tunnel coupled to $C$ with amplitude $t$.
Because of the charging effect, the total number of electrons is zero or one and the 
applied bias is very large.
Setting up the master equation as done in the main text, we obtain the
formula of steady current:
\begin{eqnarray}
I&=&\frac{2\gamma_L\gamma_R\Delta^2}{\{2\gamma_L+\gamma_R+\frac{\gamma_L}{t^2}
(2(\epsilon_0-\varepsilon_c)^2+\frac{\Delta^2+\gamma_R^2}{2})\}\Delta^2+8\gamma_Lt^2},
\end{eqnarray}
where $\epsilon_0=\frac{\varepsilon_A+\varepsilon_B}{2}$ and 
$\Delta=\varepsilon_B-\varepsilon_A$.
This formula resembles the result 
of Eq.~(\ref{eq:zeroflux}) with $\alpha_R=1$, where the current is strongly suppressed
near $\Delta=0$, while the behavior for large $|\Delta|$ is different.

\end{document}